\documentstyle[preprint,aps]{revtex}
\begin{document}
\draft
\begin{title}
{Comment on `Experimental and Theoretical Constraints of Bipolaronic
Superconductivity in High $T_{c}$ Materials: An Impossibility' }
\end{title}
\author{A.S. Alexandrov}
\address
{Department of Physics, Loughborough University, Loughborough LE11
3TU, U.K.}

\maketitle
\begin{abstract}
We show that objections raised by Chakraverty $et$ $al$ (Phys.
Rev. Lett. ${\bf 81}$, 433 (1998)) to the bipolaron model of
superconducting cuprates are the result of an incorrect approximation
for the bipolaron energy spectrum and  misuse of the bipolaron
theory. The consideration, which takes into account the multiband
energy structure of bipolarons and the unscreened electron-phonon
interaction clearly indicates that cuprates are in the Bose-Einstein
condensation regime with mobile charged bosons.
\end{abstract}
\narrowtext

Recently Chakraverty $et$ $al$ \cite{cha} raised objections to the bipolaron
model of superconducting
cuprates \cite{alemot}. They claimed that:  $a)$  Bose-Einstein
condensation
(BEC)  is not realised for any preformed pairs because  this scenario cannot
account  for the observed value of $T_{c}$, $(b)$ bipolarons are ruled
out  as superconducting
 pairs because they are too heavy, $(c)$ the
coherence length in cuprates is such that pairs are strongly overlapped,
and $(d)$ the bipolaron theory is incompatible with   photoemission
experiments.
Here I show  that  their objections   are the result of an incorrect
approximation for the bipolaron energy spectrum and misuse of our
theory.

 It is commonly accepted
that  new superconductors lie close to  BEC. We can now
 assess   how close. The bipolaron
energy spectrum has been  derived  for perovskites \cite{ale}.  It
 consists of two energy bands,
$E^{x,y}_{\bf k}=2[t\cos( k_{x,y}a)+t'\cos(
k_{y,x}a)+t_{\perp}\cos(k_{z}d)]$
where  $a$ and $d$ are the lattice constants and $t$, $t'$ ($\simeq -t/4$)
 and $t_{\perp}$ ($<<t$)  the bipolaron hopping integrals.
By  expressing the band-structure parameters
through the in-plane and out-of-plane penetration depth,
 $\lambda_{ab}=[m_{x}m_{y}c^{2}/8\pi
 n_{B}e^{2}(m_{x}+m_{y})]^{1/2}$,
 $\lambda_{c}=[m_{c}c^{2}/16\pi
 n_{B}e^{2}]^{1/2}$, one readily
obtains $T_{c}$ from the density sum rule \cite{alemot},
 \begin{equation}
k_{B}T_{c}={d c^{2}\hbar^{2}*\over{20 \lambda_{ab}^{2}e^{2} [1+\ln (16 \pi n_{B}
k_{B}T_{c}e^{2}\lambda_{c}^{2}d^{2}/\hbar^{2}c^{2})]}}.
\end{equation}
Here $m_{x}=\hbar^{2}/2ta^{2}, m_{y}=4m_{x}$, and
$m_{c}=\hbar^{2}/2|t_{\perp}|d^{2}$.
 This expression provides an unambiguous parameter-free estimate of
$T_{c}$.
In particular,
 we obtain
$T_{c}\simeq 100$ K for YBa$_{2}$Cu$_{3}$O$_{7}$ ($\lambda_{ab}=1600$ \AA
\cite{cha},  $\lambda_{c}=12600$ \AA \cite{pan} and the bipolaron
density $n_{B}=6 \times
10^{21}$ cm$^{-3}$) assuming that all $2x$ holes are bound into
bipolarons in the doped Mott insulator YBa$_{2}$Cu$_{3}$O$_{6+x}$.
 The measured value is $T_{c}=92$ K \cite{pan}. Hence,
 the consideration, which takes
 into account the multiband  energy structure
  clearly indicates  that cuprates are in the BEC
regime, while the
erroneous
approximation \cite{cha}  yields $T_{c}$ about three
times higher.

We now check to what extent our expression for the bipolaron hopping
integral, $t=0.25t_{band}e^{-g^2}$
\cite{ale} corresponds to the value of the penetration depth. Here
 $g^{2}\equiv \gamma \alpha^{2}$ with
$\gamma= \sum_{\bf q} \gamma^{2}({\bf q}) [1-\cos (q_{x}a)]/
\sum_{\bf q} \gamma^{2}({\bf q})$ and $\alpha^{2}=\epsilon_{p}/\hbar
 \omega$.
A strong  interaction of carriers with  $c$-axis
longitudinal phonons
 (the frequency $\hbar \omega \simeq 74$ meV)
 has been established
experimentally  in $YBa_{2}Cu_{3}O_{6+x}$  while
the coupling with other phonons is weak \cite{tim}.
Because of a low $c$-axis conductivity this interaction
is the  Fr\"ohlich unscreened interaction
with a ${\bf q}$-dependent matrix element, $\gamma ({\bf q}) \sim
q_{z}/q^{2}$ \cite{ale}, which yields $\gamma
=0.162$.
As a result,
one obtains  $m_{x}= 12m_{e}$ with the same polaron
level shift
 $\epsilon_{p}=250$ meV and  bare hopping
 $t_{band}=150$ meV as in Ref. \cite{cha}.
This bipolaron mass is close to $m_{x}\simeq 14 m_{e}$ in
 $YBa_{2}Cu_{3}O_{7}$ obtained
 from the experimental
$\lambda_{ab}$. It is also close to $m^{*}\simeq 13 m_{e}$ found
 in
those optical experiments \cite{mic}, which distinguish between
incoherent and Drude contributions. Mott and I always
  emphasised
  that small bipolarons in cuprates are
$intersite$ pairs \cite{alemot} as a result of $unscreened$
  interaction \cite{ale}.
By considering $onsite$
bipolarons  or by leaving out the  coefficient $\gamma$
in our expression for
  intersite  bipolaron hopping Chakraverty $et$ $al$ misuse
the theory and overestimate the small bipolaron mass by $two$ $orders$ of
magnitude.

 The coherence
length of the charged Bose-gas has nothing to do with the size of a boson.
It can be as large as in the BCS superconductor
\cite{alemot}.
 Hence, one cannot distinguish the BCS and BEC by
arguing, $(c)$, that the coherence area is  large
to accommodate many
holes. What is really conclusive is  the critical
behaviour. All experiments so far unambiguously show  that  the
upper critical field  and
 the specific heat near the
transition are  those of charged bosons \cite{alemot}.
The claim, $(d)$, that with bipolarons `one
should expect essentially ${\bf k}$-independent spectral functions
showing a ${\bf k}$-independent  gap'  \cite{cha} is not correct either.
When an electron
is emitted from the sample, the
resulting  hole propagates  in the polaron band.
It is the dispersion of this  band which is measured by ARPES.
 The normal state gap is also
${\bf k}$-dependent  owing to the polaron and bipolaron band dispersion.
In fact,  the bipolaron theory describes all major features of the
tunnelling  and photoemission spectra  \cite{ale4}.

 To the
  best of our knowledge there are no unambiguous experimental
  facts so far which are qualitatively in disagreement with the bipolaron
  theory. Of course, new  facts may.
  What is clear, however, is that any theory,
   beautiful or not,  cannot be destroyed  by `ugly' artifacts as
those in ref.\cite{cha}.

\vspace{0.5cm}

{\it A.S. Alexandrov,
Department of Physics, Loughborough University, Loughborough LE11
3TU, U.K.}

\end{document}